\documentclass[letterpaper]{article}
\usepackage{natbib}
\usepackage{alifeconf}
\usepackage{amsmath,amssymb}
\usepackage{amscd}
\usepackage[all]{xy}
\usepackage{url}

\title{Robustness and Directed Structures in Ecological Flow Networks}
\author{Taichi Haruna$^{1}$ \\
\mbox{}\\
$^1$Department of Earth \& Planetary Sciences, Graduate School of Science, Kobe University, \\
1-1, Rokkodaicho, Nada, Kobe 657-8501, Japan\\
tharuna@penguin.kobe-u.ac.jp}

\begin{document}
\maketitle

\begin{abstract}
Robustness of ecological flow networks under random failure of arcs is considered 
with respect to two different functionalities: coherence and circulation. 
In our previous work, we showed that each functionality is associated with 
a natural path notion: lateral path for the former and directed path for the latter. 
Robustness of a network is measured in terms of the size of the giant laterally connected 
arc component and that of the giant strongly connected arc component, respectively. 
We study how realistic structures of ecological flow networks affect the robustness 
with respect to each functionality. To quantify the impact of realistic network structures, 
two null models are considered for a given real ecological flow network: 
one is random networks with the same degree distribution and the other is those with 
the same average degree. Robustness of the null models is calculated by theoretically 
solving the size of giant components for the configuration model. 
We show that realistic network structures have positive effect on robustness for coherence, 
whereas they have negative effect on robustness for circulation. 
\end{abstract}

\section{Introduction}
Networks have been usually considered as undirected in the field of complex networks \citep{Newman2003}. 
However, many real-world networks are directed so that the direction of interaction 
is important for the functioning of the systems. Recently, it has been revealed that 
directed networks have richer structures such as directed assortativity \citep{Foster2010} 
and  flow hierarchy \citep{Mones2013}. 

In our previous work, we proposed a new path notion involving directedness called 
lateral path that can be seen as the dual notion to the usual directed path \citep{Haruna2011}. 
Based on category theoretic formulation, we derived the lateral path as a natural path notion associated with the dynamic mode 
of biological networks: a network is a pattern constructed by gluing functions of entities constituting the network \citep{Haruna2012}. 
Thus, its functionality is coherence, whereas the functionality of the directed path is transport. 
We showed that there is a division of labor with respect to the two functionalities within a network 
for several types of biological networks: gene regulation, neuronal and ecological ones \citep{Haruna2012}. 
It was suggested that the two complementary functionalities are realized in biological systems 
by making use of the two ways of tracing on a directed network, namely, lateral and directed. 

In this paper, we address robustness of ecological flow networks with respect to the lateral path and directed path, respectively. 
Since the natural connectedness notion associated with the directed path is the strong connectedness, we 
consider robustness of the giant strongly connected component (GSCC) for the latter. For the former, robustness of 
the giant lateral connected component (GLCC) is of interest. Thus, 
we assess robustness of ecological flow networks in terms of two different functionalities, namely, coherence and 
circulation, both of which are important for the functioning of them \citep{Ulanowicz1997}. 

Robustness of ecological networks is an intriguing issue in recent studies \citep{Montoya2006,Bascompte2009}. 
Initially, robustness of general complex networks has been argued qualitatively in terms of critical thresholds 
for the existence of the giant component \citep{Albert2000,Cohen2001}. 
For ecological networks, their robustness has been measured by the size of secondary extinctions \citep{Sole2001,Dunne2002}. 
Here, we employ a recently proposed idea to measure robustness quantitatively \citep{Schneider2011,Herrmann2011}. 
As a first step, we consider only random failure of arcs. The size of giant components is measured by 
the number of arcs involved because laterally connected components are defined only on the set of arcs. 

Here, we study the impact of realistic network structures on robustness with respect to the two functionalities. 
Two complementary measures of it are proposed by comparing the robustness of a given real network with that of 
the two null models: random networks with the same degree distribution and those with the same average degree. 
The robustness of the two null models is calculated by theoretically solving the percolation problem on 
the configuration model, random networks with an arbitrary degree distribution \citep{Newman2001}. 

This paper is organized as follows. 
In Section 2, we develop a theory to calculate the size of GLCC and GSCC under 
random removal of arcs in the configuration model. In Section 3, we propose 
two measure for the impact of realistic structures on robustness of networks 
by using the theoretical result obtained in Section 2. In Section 4, the proposed 
measures are applied to 10 ecological flow networks. In Section 5, we discuss 
the results and indicate future directions. 

\section{Random Removal of Arcs in the Configuration Model}
In this section, we consider a percolation problem, 
random removal of arcs, in the configuration model 
with respect to the lateral connectedness and the strong connectedness. 

\begin{figure}[t]
\begin{center}
\includegraphics[width=7cm]{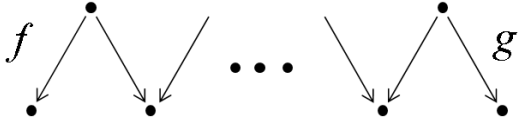}
\caption{An example of lateral path.}
\label{fig1}
\end{center}
\end{figure}

A \textit{lateral path} in a directed network is a path in 
the network such that the direction of arcs involved changes alternately \citep{Haruna2012} (Fig.~\ref{fig1}). 
Two arcs are called \textit{laterally connected} if they are connected by 
a lateral path \citep{Haruna2011}. Lateral connectedness defines an equivalence relation 
on the set of arcs. Each equivalence class is called \textit{laterally connected 
component}. 

Since lateral connectedness is defined on the set of arcs, here we also consider 
strong connectedness for arcs. Two arcs are called \textit{strongly connected} 
if there is a directed path from one arc to the other arc, and vice versa. 

Let us consider a random directed network with degree distribution $P(k_i,k_o)$. 
$P(k_i,k_o)$ is the fraction of nodes in the network with in-degree $k_i$ and 
out-degree $k_o$. We make use of the generating function formalism 
\citep{Callaway2000,Newman2001} to calculate the sizes of giant laterally or 
strongly connected components (in short, GLCC or GSCC, respectively) after removing arcs uniformly 
at random with probability $1-\phi$, where $\phi$ is the occupation probability. 

The generating function for $P(k_i,k_o)$ is 
\begin{equation}
G(x,y)=\sum_{k_i,k_o} P(k_i,k_o)x^{k_i}y^{k_o}. 
\label{eq1}
\end{equation}
The average degree $z:=\langle k_i \rangle=\langle k_o \rangle$ is given by 
\begin{equation}
z=\frac{\partial G}{\partial x}(1,1)=\frac{\partial G}{\partial y}(1,1). 
\label{eq2}
\end{equation}
Let $P_{i}(k_i):=\sum_{k_o} P(k_i,k_o)$ be the in-degree distribution and 
$P_{o}(k_o):=\sum_{k_i} P(k_i,k_o)$ the out-degree distribution. Their 
generating functions are 
\begin{equation}
F_0(x):=G(x,1) \textrm{ and } H_0(y):=G(1,y), 
\label{eq3}
\end{equation}
respectively. 

We introduce four excess degree distributions and corresponding 
generating functions that are necessary for the calculation in what follows. 

\begin{figure}[t]
\begin{center}
\includegraphics[width=7cm]{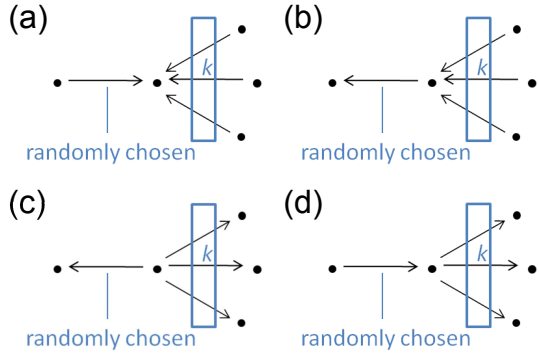}
\caption{Four excess degree distributions. See the main text for details.}
\label{fig2}
\end{center}
\end{figure}

First, let $P_0(k)$ be the probability that the number of the other arcs 
arriving at the target node of a randomly chosen arc is $k$ (Fig.~\ref{fig2} (a)). It is 
given by 
\begin{equation}
P_0(k):=\frac{1}{z}\sum_{k_0} (k+1)P(k+1,k_0)
\label{eq4}
\end{equation}
and its generating function is 
\begin{equation}
F_{1,0}(x):=\sum_{k} P_0(k)x^k=\frac{1}{z}\frac{\partial G}{\partial x}(x,1)=\frac{1}{z}\frac{\partial F_0}{\partial x}(x). 
\label{eq5}
\end{equation}

Second, let $P_1(k)$ be the probability that the number of arcs 
arriving at the source node of a randomly chosen arc is $k$ (Fig.~\ref{fig2} (b)). It is 
given by 
\begin{equation}
P_1(k):=\frac{1}{z}\sum_{k_0} k_0 P(k,k_0)
\label{eq6}
\end{equation}
and its generating function is 
\begin{equation}
F_{1,1}(x):=\sum_{k} P_1(k)x^k=\frac{1}{z}\frac{\partial G}{\partial y}(x,1). 
\label{eq7}
\end{equation}

Third, let $Q_0(k)$ be the probability that the number of the other arcs 
leaving from the source node of a randomly chosen arc is $k$ (Fig.~\ref{fig2} (c)). It is 
given by 
\begin{equation}
Q_0(k):=\frac{1}{z}\sum_{k_i} (k+1)P(k_i,k+1)
\label{eq8}
\end{equation}
and its generating function is 
\begin{equation}
H_{1,0}(y):=\sum_{k} Q_0(k)y^k=\frac{1}{z}\frac{\partial G}{\partial y}(1,y)=\frac{1}{z}\frac{\partial H_0}{\partial y}(y). 
\label{eq9}
\end{equation}

Finally, let $Q_1(k)$ be the probability that the number of arcs 
leaving from the target node of a randomly chosen arc is $k$ (Fig.~\ref{fig2} (d)). It is 
given by 
\begin{equation}
Q_1(k):=\frac{1}{z}\sum_{k_i} k_i P(k_i,k)
\label{eq10}
\end{equation}
and its generating function is 
\begin{equation}
H_{1,1}(y):=\sum_{k} Q_1(k)y^k=\frac{1}{z}\frac{\partial G}{\partial x}(1,y). 
\label{eq11}
\end{equation}

\subsection{Giant Laterally Connected Component}
Let $u$ be the average probability that an arc is not connected to 
the GLCC via a particular arc with the same target and $v$ 
the average probability that an arc is not connected to 
the GLCC via a particular arc with the same source. 
Then, the average probability that an occupied arc does not 
belong to the GLCC is 
\begin{equation}
\sum_{k,l} P_0(k) u^k Q_0(l) v^l=F_{1,0}(u)H_{1,0}(v). 
\label{eq12}
\end{equation}
Hence, the size of the GLCC is 
\begin{equation}
L=\phi(1-F_{1,0}(u)H_{1,0}(v)).
\label{eq13}
\end{equation}
The values of $u$ and $v$ are calculated by the following set of equations: 
\begin{equation}
\begin{cases}
u=\sum_{k}Q_0(k)(1-\phi + \phi v^k)=(1-\phi)+\phi H_{1,0}(v) \\
v=\sum_{k}P_0(k)(1-\phi + \phi u^k)=(1-\phi)+\phi F_{1,0}(u). 
\end{cases}
\label{eq14}
\end{equation}

The critical occupation probability $\phi_{L,c}$ for the appearance 
of GLCC can be obtained from the linear stability analysis of the 
trivial solution $(u,v)=(1,1)$ of (\ref{eq14}). It turns out to be 
\begin{equation}
\phi_{L,c}=\frac{z}{\sqrt{\left( \langle k_i^2 \rangle - z \right) \left( \langle k_o^2 \rangle - z \right)}}. 
\label{eq15}
\end{equation}

\subsection{Giant Strongly Connected Component}
The calculation of the size of the GSCC is similar to 
the node component case \citep{Dorogovtsev2001,Schwartz2002}. 
In \citep{Serrano2007}, five notions of edge components are 
considered. For our purpose, consideration on the 
usual three components (in-, out- and strongly connected) 
as in the node component case are enough. However, these are 
implicit in the following calculation. 

Let $u$ be the average probability that an arc is not connected to 
the GSCC via a particular arc leaving from its target and $v$ 
the average probability that an arc is not connected to 
the GSCC via a particular arc arriving at its source. 
Then, the average probability that an occupied arc does 
belong to the GSCC is 
\begin{equation}
\sum_{k,l} Q_1(k)(1-u^k) P_1(l)(1-v^l)=(1-H_{1,1}(u))(1-F_{1,1}(v)). 
\label{eq16}
\end{equation}
Hence, the size of the GSCC is 
\begin{equation}
S=\phi(1-H_{1,1}(u))(1-F_{1,1}(v)).
\label{eq17}
\end{equation}
The values of $u$ and $v$ are calculated by the following set of equations: 
\begin{equation}
\begin{cases}
u=\sum_{k}Q_1(k)(1-\phi + \phi u^k)=(1-\phi)+\phi H_{1,1}(u) \\
v=\sum_{k}P_1(k)(1-\phi + \phi v^k)=(1-\phi)+\phi F_{1,1}(v). 
\end{cases}
\label{eq18}
\end{equation}

The critical occupation probability $\phi_{S,c}$ for the appearance 
of GSCC is given by 
\begin{equation}
\phi_{S,c}=\frac{z}{\langle k_i k_o \rangle}, 
\label{eq19}
\end{equation}
which is the same as in the node component case \citep{Schwartz2002}. 

\begin{figure}[!htb]
\begin{center}
\includegraphics[width=7cm]{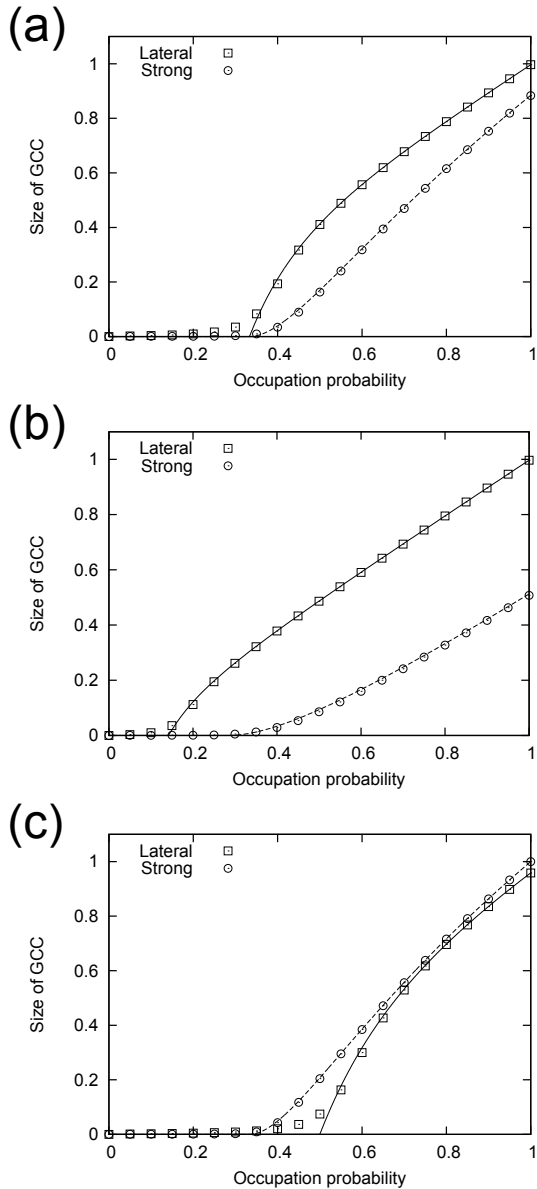}
\caption{$L(\phi)$ and $S(\phi)$ for (a) the uncorrelated Poisson distribution with $\lambda=3$, 
(b) the uncorrelated exponential distribution with $\kappa=4$ and 
(c) the correlated Poisson distribution with $\lambda=2$. Lines are theoretically obtained. For (a) and (c), 
(\ref{eq14}) and (\ref{eq18}) are numerically solved. For (b), we obtain analytic expressions. 
Squares and circles are numerical simulations and averaged over 1000 different random removal sequences 
on different configuration model networks with the number of nodes 500 for (a) and (b), and 
1000 for (c).}
\label{fig3}
\end{center}
\end{figure}

\subsection{Examples}
We calculate the sizes of the GLCC and the GSCC as functions of the occupation probability $\phi$ 
for three degree distributions: 
(a) Uncorrelated Poisson distribution (UPD) 
\begin{equation}
P(k_i,k_o)=\frac{e^{-2\lambda}\lambda^{k_i+k_o}}{k_i ! k_o !}, 
\label{eq20}
\end{equation}
(b) Uncorrelated exponential distribution (UED) 
\begin{equation}
P(k_i,k_o)=\left( 1 - e^{-1/\kappa}\right)^2 e^{-\frac{k_i+k_o}{\kappa}}, 
\label{eq21}
\end{equation}
and (c) Correlated Poisson distribution (CPD) 
\begin{equation}
P(k_i,k_o)=\frac{e^{-\lambda}\lambda^{k_i}}{k_i !} \delta_{k_i,k_o}, 
\label{eq22}
\end{equation}
where $\lambda,\kappa>0$ are parameters and $\delta_{k_i,k_o}$ is the Kronecker delta. 
The results are compared with numerical simulations in Fig.~\ref{fig3}, which shows 
that the agreement between simulation and theory is well. 

For critical occupation probabilities, we have 
$\phi_{L,c}=\phi_{S,c}=1/\lambda$ for UPD, 
$\phi_{L,c}=\left( e^{1/\kappa}-1 \right)/2 < \left( e^{1/\kappa}-1 \right)=\phi_{S,c}$ for UED 
and $\phi_{L,c}=1/\lambda > 1/(\lambda+1)=\phi_{S,c}$ for CPD. 
Thus, these examples also show that all possibilities 
$\phi_{L,c}=\phi_{S,c}$, $\phi_{L,c}>\phi_{S,c}$ and $\phi_{L,c}<\phi_{S,c}$ 
actually occur.

\section{Two Measures for Impact of Realistic Structures on Robustness}

\subsection{Robustness}
Given a directed network, let $L(\phi)$ be the size of the GLCC 
and $S(\phi)$ the size of the GSCC for occupation probability $\phi$. 
Motivated by the robustness measure proposed in \citep{Schneider2011,Herrmann2011}, 
we define the robustness of the GLCC and that of the GSCC by 
\begin{equation}
R_L=\int_0^1 L(\phi) d\phi \textrm{ and } R_S=\int_0^1 S(\phi) d\phi, 
\label{eq23}
\end{equation}
respectively. 

Our robustness measure is similar to link robustness in \citep{Zeng2012}, 
however, since we measure the size of a component by the number of arcs 
belonging to it, it is different from link robustness. In particular, since 
$L(\phi)$ and $S(\phi)$ cannot exceed the diagonal line, we have 
$R_L, R_S \leq 0.5$.  

\subsection{Gain}
Given a directed network, we would like to consider how much its robustness (of the GLCC or the GSCC) 
is enhanced or degraded compared to a reference network. One measure is the ratio of the robustness of the 
given network to that of the reference network \citep{Schneider2011}. 
We call this measure \textit{robustness gain}. If we denote the robustness of the given network by $R_{given}$ 
and that of the reference network by $R_{ref}$, then the robustness gain is defined by 
\begin{equation}
G_{given/ref}:=R_{given}/R_{ref}. 
\label{eq24}
\end{equation}
We here consider three combinations of given-reference pairs: 
(given,ref)=(real, config), (given,ref)=(config, Poisson) and (given,ref)=(real,Poisson), 
where `real' indicates a real-world network, `config' the configuration model network with the same degree distribution and 
`Poisson' the (uncorrelated) Poissonian network with the same average degree. 
The robustness gains for the three given-reference pairs are denoted by 
$G_{r/c}, G_{c/p}$ and $G_{r/p}$, respectively. Note that $G_{r/p}=G_{r/c}G_{c/p}$. 

\subsection{Complement Ratio}
The other way to measure the effect of realistic structures on robustness is to 
evaluate the amount of unrealized robustness of the reference network 
(namely, $0.5-R$) utilized by the given network. We define the 
\textit{robustness complement ratio} for the above three combinations of given-reference 
pairs by 
\begin{equation}
C_{given/ref}:=\frac{R_{given}-R_{ref}}{0.5-R_{ref}}, 
\label{eq25}
\end{equation}
where $(given,ref)=(r,c)$, $(c,p)$ or $(r,p)$. 

Both $G_{given/ref}$ and $C_{given/ref}$ are considered for the lateral connectedness 
and the strong connectedness in next section. We write $G_{L,given/ref}$ and $C_{L,given/ref}$ for 
the former and $G_{S,given/ref}$ and $C_{S,given/ref}$ for the latter. 

\section{Ecological Flow Networks}
In this section, we apply the indexes introduced in previous section to 
relatively large 10 networks (with the number of arcs $>100$) among 48 
flow networks collected by R. Ulanowicz. Data are downloaded from 
\url{http://www.cbl.umces.edu/~ulan/ntwk/network.html}. 

\begin{figure}[!htb]
\begin{center}
\includegraphics[width=7cm]{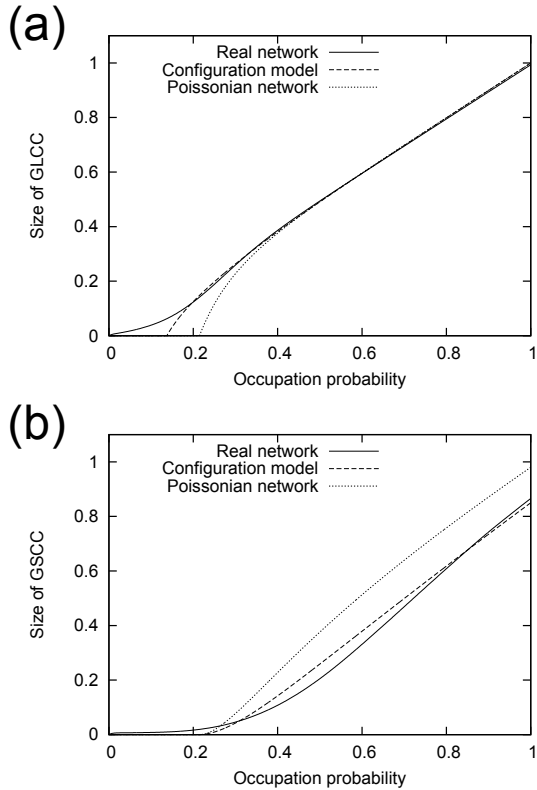}
\caption{(a) $L(\phi)$ and (b) $S(\phi)$ for (vii) Middle Chesapeake Bay in Summer network (solid lines), 
those for the configuration model network with the same degree distribution (dashed lines) 
and those for the Poissonian network with the same average degree (dotted lines). 
}
\label{fig4}
\end{center}
\end{figure}

\begin{figure}[!htb]
\begin{center}
\includegraphics[width=7cm]{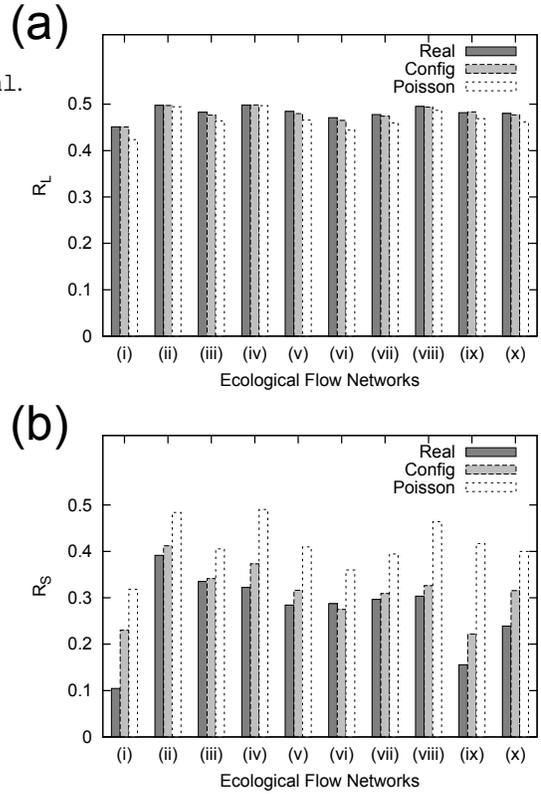}
\caption{Robustness of (a) the GLCC and (b) the GSCC for the 10 ecological flow networks. 
Real: original networks, Config: the configuration model networks with the same degree 
distribution and Poisson: the Poissonian networks with the same average degree.}
\label{fig5}
\end{center}
\end{figure}

\begin{figure}[!htb]
\begin{center}
\includegraphics[width=7cm]{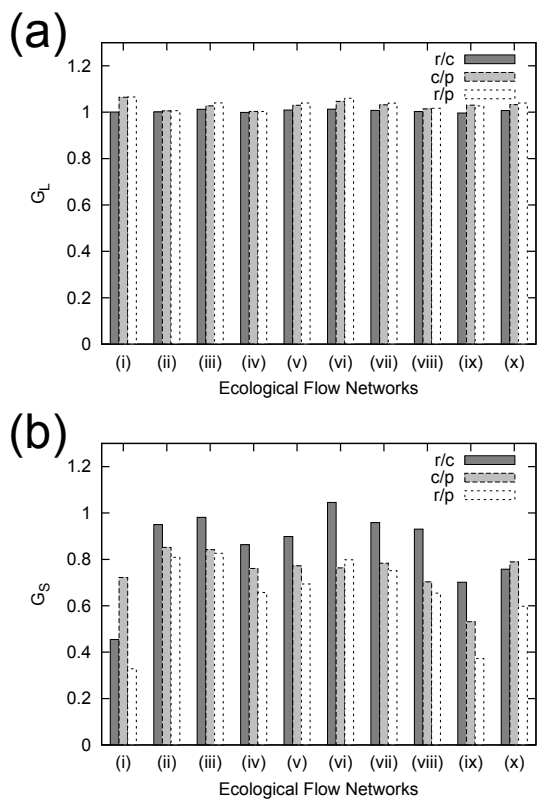}
\caption{Robustness gain of the 10 ecological flow networks for (a) the GLCC and (b) the GSCC. 
Three given and reference network pairs are considered. 
r/c: (given,ref)=(real,config), c/p: (given,ref)=(config,Poisson) and r/p: (given,ref)=(real,Poisson). 
See the main text for details.}
\label{fig6}
\end{center}
\end{figure}

\begin{figure}[!htb]
\begin{center}
\includegraphics[width=7cm]{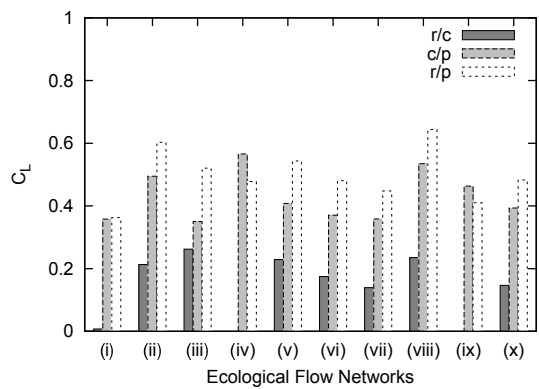}
\caption{Robustness complement ratio of the 10 ecological flow networks for the GLCC. 
Three given-reference network pairs are considered. 
r/c: (given,ref)=(real,config), c/p: (given,ref)=(config,Poisson) and r/p: (given,ref)=(real,Poisson). 
Data that have negative values are omitted. $C_S$ has negative values except one case (data not shown). 
See the main text for details.}
\label{fig7}
\end{center}
\end{figure}

\subsection{Data}
Here, we list the 10 ecological flow networks we analyze. 
In the following, $N$ is the number of nodes and $A$ is the number of arcs 
included in the largest weakly connected component. 
$z=\langle k_i \rangle=\langle k_o \rangle$ is the average degree. 
The number associated to each network is the web number in the original data source. 
In every network, each arc indicates the existence of carbon flow from 
its source to target. 
(i) Chesapeake Bay Mesohaline Network ($N=26, A=122, z=3.4$, Web 34). 
(ii) Everglades Graminoids Wet Season ($N=66, A=793, z=12.0$, Web 40). 
(iii) Final Narragansett Bay Model ($N=32, A=158, z=4.9$, Web 42). 
(iv) Florida Bay Wet Season ($N=125, A=1938, z=15.5$, Web 38). 
(v) Lake Michigan Control Network ($N=34, A=172, z=5.1$, Web 47). 
(vi) Lower Chesapeake Bay in Summer ($N=29, A=115, z=4.0$, Web 46). 
(vii) Middle Chesapeake Bay in Summer ($N=32, A=149, z=4.7$, Web 45). 
(viii) Mondego Estuary - Zostrea Site ($N=43, A=348, z=8.1$, Web 41). 
(ix) St Marks River (Florida) Estuary ($N=51, A=270, z=5.3$, Web 43). 
(x) Upper Chesapeake Bay in Summer ($N=33, A=158, z=4.8$, Web 44). 

\subsection{Results}
We plot $L(\phi)$ (Fig.~\ref{fig4} (a)) and $S(\phi)$ (Fig.~\ref{fig4} (b)) for 
(vii) Middle Chesapeake Bay in Summer network, 
the configuration model network with the same degree distribution and the Poissonian network 
with the same average degree, as a typical example. 
$L(\phi)$ and $S(\phi)$ for real ecological flow networks are calculated 
by averaging the size of the largest connected components over 1000 random removal 
sequences of arcs. 

The robustness values for all 10 networks are shown in Fig.~\ref{fig5}. 
One can see opposite tendency on how realistic structures influence robustness 
between the GLCC and the GSCC. $R_L$ tends to increase as more realistic 
structures are imposed on one hand, $R_S$ tends to decrease on the other hand. 
However, since $R_L$ is close to 0.5 already for the Poissonian network in most 
cases, the robustness gain for the GLCC is almost unity in all three 
given-reference pairs as seen in Fig.~\ref{fig6} (a). 
For $R_S$, one can see that the realistic degree distributions are the dominant factor for 
the degradation of robustness in most cases from Fig.~\ref{fig6} (b). 

The tendency that realistic structures have positive impact on robustness of 
the GLCC can be captured more clearly by the robustness complement ratio 
as shown in Fig.~\ref{fig7}. One can also see that the realistic degree distributions 
are the dominant factor to enhance the robustness of the GLCC in most cases. 

\section{Discussions}
Whether realistic structures of ecological networks have positive impact on 
their robustness or stability or not is controversial \citep{Allesina2012}. 
The answer to this question generally depends on the types of ecological 
interaction and dynamic processes of interest \citep{Thebault2010,Allesina2012}. 
In this paper, we focused on robustness of ecological flow networks under 
random failure of arcs with respect to the two different functionalities, namely, 
coherence and circulation. The former is captured by the robustness of the 
GLCC and the latter by that of the GSCC. We found that they exhibit 
opposite tendency for constraints by the realistic network structures: 
the realistic network structures enhance the robustness of the GLCC on one hand, 
they degrade that of the GSCC on the other hand. In both case, it is 
suggested that the realistic degree distributions are one of the most important factors. 

The former result seems to be consistent with the food-web stabilizing factor 
proposed in \citep{Gross2009}: 
``(i) species at high trophic levels feed on multiple prey species and 
(ii) species at intermediate trophic levels are fed upon by multiple predator species'', 
because such patterns in a network could contribute to make multiple lateral paths between arcs. 
Whereas, the latter result could provide a quantitative support for the 
`autocatalytic view' on ecological flow networks proposed by R. Ulanowicz 
\citep{Ulanowicz1997}. 

Our result in this paper suggests that complex networks can be both robust and fragile 
in a different sense from that in \citep{Albert2000}: under the same attack strategy, 
robust for one functionality and fragile for another functionality. 

It is of interest whether the same tendency can be seen or not for the other 
various attack strategies \citep{Holme2002} and for the other 
kinds of directed biological networks such as gene regulation and brain. 
Research results on these issues will be reported elsewhere near future. 

\section{Acknowledgements}
This work was partially supported by JST PRESTO program. 

\footnotesize
\bibliographystyle{apalike}
\bibliography{ecal13}

\end{document}